# Survey on Error Concealment Strategies and Subjective Testing of 3D Videos

Md. Mehedi Hasan, Michael Frater, and John F. Arnold
School of Engineering and Information Technology
University of New South Wales,
UNSW Canberra, Canberra, Australia
Md.Hasan@student.adfa.edu.au, M.Frater@adfa.edu.au, J.Arnold@adfa.edu.au

*Abstract*— Over the last decade, different technologies to visualize 3D scenes have been introduced and improved. These technologies include stereoscopic, multi-view, integral imaging and holographic types. Despite increasing consumer interest; poor image quality, crosstalk's or side effects of 3D displays and also the lack of defined broadcast standards has hampered the advancement of 3D displays to the mass consumer market. Also, in real time transmission of 3DTV sequences over packet-based networks may results in visual quality degradations due to packet loss and others. In the conventional 2D videos different extrapolation and directional interpolation strategies have been used for concealing the missing blocks but in 3D, it is still an emerging field of research. Few studies have been carried out to define the assessment methods of stereoscopic images and videos. But through industrial and commercial perspective, subjective quality evaluation is the most direct way to evaluate human perception on 3DTV systems. This paper reviews the state-of-the-art error concealment strategies and the subjective evaluation of 3D videos and proposes a low complexity frame loss concealment method for the video decoder. Subjective testing on prominent datasets videos and comparison with existing concealment methods show that the proposed method is very much efficient to conceal errors of stereoscopic videos in terms of computation time, comfort and distortion.

## I. Introduction

With the rapid development of 3DTV technologies and videos, influenced the investigation of 3D video Quality of Experience (QoE) [1] in its processing chain, especially in the coding, transmission and display stages. A typical video processing chain (dataflow) may be divided into several steps and different artefacts can be created at the time of this processing chain as shown in Fig. 1. The first step is the creation and capture step. 3D Videos can be captured by two or more synchronized cameras in a stereoscopic view or, multiview setup and also can be augmented by depth information captured with a special sensor. 3D contents can also be rendered from a 3D model of the scene using different computer graphics drawing techniques. Unnatural correspondences, miss synchronization and conversion between two stereo images or videos lead to several disturbing artifacts. In the multiview representation, each view is represented as an individual image while in the image & depth representation each scene is represented by an image and a depth map from which the individual views can be rendered. Each of these representations may cause specific artifacts such as vertical parallax for multiview and disocclusions [2]. Typical image and video compression schemes for 2D and 3D [3] exploit the spatial, temporal redundancies (lossless) and irrelevancies (lossy) to compress the source signal. This may lead to 2D artifacts such as blocking, blurring, and ringing, which may also influence the 3D perception. One of the most common problems in digital transmission is packet loss, which may influence the image and video quality considerably. Most of the error resilience and concealment techniques have been designed for 2D video, but they may influence the 3D quality as well. Various technologies have been proposed to deliver stereoscopic videos through partitioned optical channels to the viewer's eyes, including anaglyph, polarization, shutter and auto-stereoscopic [4]. Each of these approaches has its specific artifacts which are also highly scene dependent.

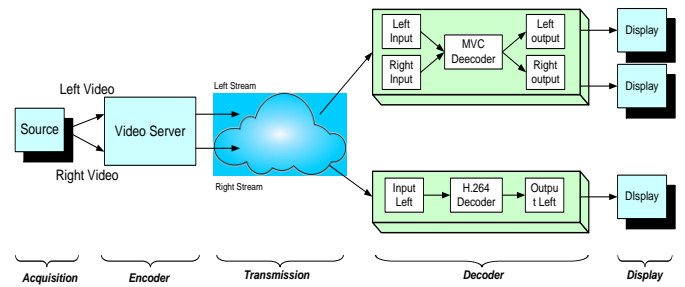

Fig. 1. Overview of 3D Video Processing Chain.

This paper brings down all the recent trends of error concealment of 3D videos with the subjective testing scenarios. Various analysis has been performed with our proposed method, which is a low complexity full frame loss concealment and feasible for real-time video applications.

The rest of the paper is organized as follows: In section II, the transmission errors with its error resilience and concealment have been discussed. In Section III, different kind of concealment strategies for different type of 3D videos have be analyzed. Subjective testing of these methods is presented in IV. Our proposed method and its experimental results are discussed in V and VI, before concluding the work in section VII.

## II. BACKGROUND STUDIES

### A. Trnasmission or Broadcasting artifacts

Apart from the artifacts introduced by different video processing steps, the transmission network itself often affected by errors due to delay or packet loss. H.264/AVC [5] has a spatial-temporal block-based structure. The propagated bit-streams are very responsive to transmission errors leading to spatial and temporal error propagation. Spatial error propagation occurs when there is a loss of synchronization in predictive or entropy decoding. Also, Multiview Video Coding (MVC) [6] has higher coding efficiency which exploits the inherent redundancy by interview prediction. It assumes that the lower network layer has the capability to detect and drop corrupted packets so that the decoder is only presented with intact packets [7]. Therefore, error resilient coding is required to limit the propagation of visual artifacts while error concealment can be adopted to minimize the visual artifacts caused by the lost slices [8].

The impacts of transmission of video network errors on 3D video quality have been discussed in many studies. The subjective experiment results from [9] showed that depending on the scenarios the same packet loss has a significantly different impact on user perceived video quality. The relationship between bit rates and perceptual quality of HDTV contents was investigated in [10], where the results showed that a small difference in bit rates is linked to a large difference in quality. The experimental results from [11] illustrated that users were more affected and annoyed by long and widely spread packet losses than bursts. When information is lost, e.g. because of dropped packets, error concealment methods are often used at the end user side to reconstruct the error affected signal. In Fig. 2 in video transmission chain, error resilience and concealment are introduced in the provider side and user side respectively.

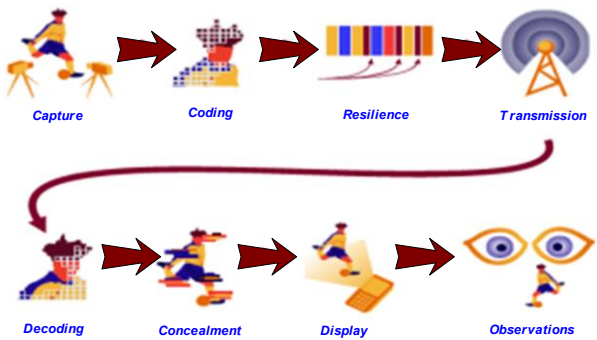

Fig. 2. Error resilience and concealment-based transmission.

### B. Error Resilience

Real-time video data streamed over unreliable IP networks, such as the Internet, may encounter losses due to dropped packets or late arrivals. In order to handle these errors, there exist error resilient techniques, which are performed by the sender and error concealment schemes, performed by the receiver. Error resilience coding is required for transmission system over lossy networks since video encoding uses predictions, losses propagation. In 3D video, prediction structures are more complex, and protection is required more to overcome loss propagation. There are two approaches for loss protection [12]. One solution is to use a feedback channel for the lost packets and retransmit lost packets. Feedback channel might not be available in some solutions like broadcasting. Feedback solution also cause delays and not suitable for real-time communications. Forward error correction (FEC) is another solution where data is sent with redundant error correcting codes in order to overcome losses. Multiple Description Coding (MDC) [13] is also another alternative which uses multiple descriptions of the video, where each description can be decoded independently and all of them can have a better decoding quality.

### C. Error Concealment

Apart from the artefacts introduced by the source coding, the transmission network in itself often introduces errors e.g. delay or packet loss. Error concealment methods are used at the decoder to measure the lost information by taking into account the inherent correlation of the missing blocks or pixels with its neighbouring blocks or pixels. The impacts of network errors and its concealment on 2D videos [14] [15] have been discussed in many studies. But for 3D videos [16] [17], distortion in one view or in both views is perceived differently. A degradation in one view or a temporal misalignment between the left and the right view leads to binocular rivalry. This strongly degrades the quality of experience as it exhibits visual discomfort which might lead to headache or nausea [18]. As the perception of 3D videos is different from that for 2D videos, the influence of network errors and its concealment will be more complex considering users experience and perception.

## III. ERROR CONCEALMENT OF 3D VIDEOS

### A. Error Concealment Strategies

Several techniques have been applied to conceal the error frames occurred due to transmission errors. Most of them are in pixel or, block domain and don't consider the real-time video data streaming which require less space and time complexity. Pang et al. [19] proposed an EC algorithm to conceal frame losses in stereo videos, which employs the motion vector extrapolation or the disparity vector replication. Knorr et al. [20] proposed error concealment (EC) algorithm for block losses which detects feature points around a lost block in a view, matches them to the matching points in the other view, estimates a projective mapping based on the matched pairs, and uses the mapping to fill in the lost block. Although such methods are useful for the re-construction or estimation of a portion of the image, they are not practical for full frame losses. Guenther et al. [21] also proposed an error concealment algorithm for block losses in stereoscopic images and videos, which conceals an erroneous block with a motion-compensated block or a disparity-compensated block based on the side match criterion. Clemens et al. [22] extended the maximally smooth recovery method [23], which was proposed for mono image concealment. They also used a projective mapping to exploit inter-view correlations. Xiang et al. [24] proposed a hybrid EC algorithm, which selects the best replacing block among motion-compensated blocks, disparity-compensated blocks, or their overlapped block in order to reconstruct a lost block. However, errors in multi-view video sequences can be more effectively concealed than those in stereo video sequences, since the information in more than two neighbouring views can be used for the concealment. Though H.264 has its own concealment techniques, several concealment strategies [5] [17] [25] [26] have been applied for subjective testing of packet losses in video transmission or broadcasting environment. For Among them some have been used rapidly for their good performance in 3D subjective quality evaluation. This evaluation study was conducted with the objective of finding out relevant relationships between perceive 3D content quality, packet loss statistics and frame loss concealment. Below we have shown some concealment strategies have been done so far so subjective testing.

H.264 Concealment [8] are the most computationally intensive algorithm as it involves the sophisticated error concealment implemented in the H.264 software which uses spatial and spatial-temporal interpolation depending on the frame type. However, in the 3D case, only a single view is distorted, and thus binocular rivalry may occur as the error concealment artifacts are visible only in one view. Switching to 2D [5] is another approach where video is switched from 3D to a 2D presentation when an error occurs in one view. As the other view is undistorted in our setup, this undistorted view is displayed to both eyes thus leading to a 2D impression without disparity. Frame Freeze [25] consists of copying of the last received frame. The last frame that was correctly received for only one error views is displayed while the effects of the transmission errors are affecting one view. In Double Freeze [25], last frame that was correctly received for both views is displayed while the effects of the transmission errors are affecting one view or both views. Finally, in Reduced Playback Speed [5] instead of stopping the video completely, it is assumed that a buffer of video frames exists which contains half a second of decoded content, corresponding to 12 frames. These 12 frames are slowly played back during the recovery time of the decoder. The observer would thus see that the playback slows down, skips and then continues at normal speed.

### B. Full Frame Loss Concealment of Stereoscopic Videos

Most of the error concealment techniques found in the literature can handle macroblock or slice losses using neighbouring macroblock information like motion vectors and pixel values [27]. However, in low bitrate video coding, packet losses might result in loss of a whole frame. In frame loss cases, temporal and spatial interpolation or motion extrapolation within a frame do not work. Several algorithms are proposed to conceal full frame losses in monoscopic videos but very few studies are found in the literature for stereoscopic error concealment. Earlier, we have discussed some of the existing error concealment methods for 3D videos. Where, erroneous blocks are concealed using the correspondences between the two views in the stereoscopic image pair. But they are not practical for full frame losses. One way to conceal the lost frame is to use view interpolation [28]. Bilen et. al. proposed [29] another method for full frame concealment where, following a number of steps, different algorithms are used to estimate the MVs and DVs for each MB from the source frame. Then, a median filter is applied to filter both the MV and DV fields, to fill the empty spaces and to filter irregularities. These vectors are used with motion and disparity compensation to form the concealed frame as shown in Fig. 3. Finally, the resulting picture is filtered again using a median filter to fill the empty regions. Hewage et. al. also proposed another frame concealment algorithm [30] where he correlated the colour and depth map of image sequences. The colour motion information is reused for prediction during depth map coding. The redundant motion information is then used to conceal transmission errors at the decoder. Bilen et. al. [31] also proposed motion and disparity aided error concealment of an entire frame loss.

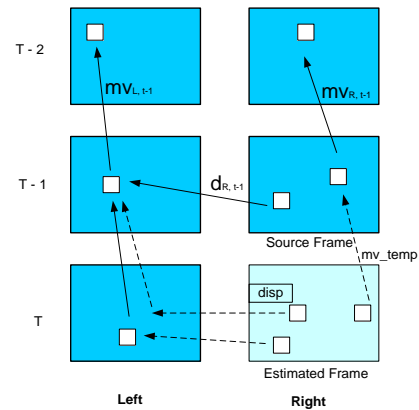

Fig. 3. Algorithm for the estimation of the right frame at time T.

### C. Concealment based on Lfet/Right View

Error concealment is a major kind of technique that effectively deals with video delivery via error-prone networks. It resides at the decoder side to fill up the lost video contents. Error concealment techniques in the traditional single view coding have been widely exploited. The simplest way is temporal replacement (TR) [33], which utilizes the zero-motion vector (MV) to reconstruct a lost macroblock (MB). Then the block matching algorithm (BMA) is proposed to select an optimal MV to substitute for the lost one [15]. In [34], a technique which combines the overlapped motion compensation and the side match criterion makes the effect of lost motion vectors subjectively imperceptible. These techniques are designed for the single-view video coding, which only considers the temporal or spatial correlations. So there will be inadequacy if they are directly applied in stereoscopic video coding without considering interview correlations. Few works [20] have been reported on the error concealment of stereoscopic video coding. How to effectively take error concealment in a DCP-based stereoscopic video coding system is still an unanswered question. In [24], Xiang et. al. proposed a novel error concealment method, based on overlapped block motion and disparity compensation (OBMDC), whose weights are determined by the side match criterion and viewpoints. Also point out that the concealed blocks of the left view should be utilized to improve the performance of the right view. So, the popular stereoscopic or multi-view video coding systems adopt both the disparity compensation prediction (DCP) [32] in the interview direction and motion compensation prediction (MCP) in the temporal direction of a single view. The basic prediction structure of the two-view based stereoscopic video coding is shown in Fig. 4.

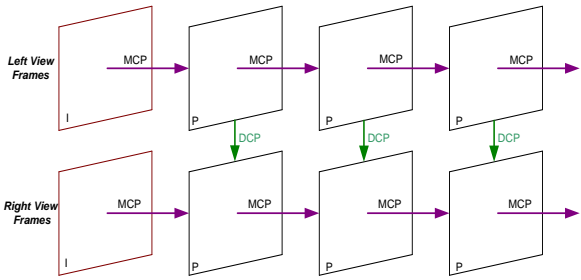

Fig. 4. Basic prediction structure of stereoscopic video.

### D. Concealment based on Depth of View

Stereoscopic video renders two slightly different views of a scene for each eye to enable the perception of depth. Recently, colour and depth map-based videos has been extensively used in research and standardization activities [35]. Monoscopic video together with its associated each pixel depth map can be used to scale existing videos with low overheads. This format is more flexible and adaptable compared to storage and transmission of the left and right views at low bitrate. A specialized image warping technique known as the Depth-Image-Based Rendering (DIBR) is used to synthesize the desired binocular viewpoint image [36]. As the motion of colour images and corresponding depth information is highly correlated, the colour motion information can thus be used as candidate motion information for the depth video. Hence, the bandwidth for low bitrate stereoscopic video applications can be reduced as motion vectors (MVs) are transmitted only once. The analysis of motion correlation of 'colour and depth' sequences and a MV sharing scheme based on MPEG-2 and H.264/AVC is described in [37]. These developments are mainly focused on reducing the bandwidth, complexity and encoding time of the stereoscopic video encoder. In [38], the authors propose a frame concealment method using shared MVs between the color and depth data. If the color frame is received corrupted, the MVs from the corresponding uncorrupted depth frame are used to form the concealed frame, and vice versa. If both corresponding frames are lost, then conventional single-view concealment algorithms are used to recover the lost frame. Also, Depth-image based temporal error concealment has been proposed by Liu et. al. [39]. Authors consider the correlation between the colour video and the depth video to propose a temporal error concealment technique for the lost MB.

### E. Concealment of Multiview Videos

Various algorithms have been proposed so far to conceal the multiview videos. As multiview videos has more information than stereoscopic videos, so concealing the frames of multiview videos is easier than stereoscopic videos. A typical multi-view sequence exhibits high spatio-temporal correlations within each view. To exploit these correlations, Merkle et al. [40] proposed the extended hierarchical B prediction structure, which is illustrated in Fig. 5. In Multiview coding, the hierarchical B prediction mode is extended so that a frame can be predicted from inter-view frames as well as intra-view frames. From the observations of the above figure, a view is referred to as I-view, P-view or B-view according to the type of its first key frame in the GOP. The first view $V_0$ is encoded using only the temporal prediction. The other even views $V_2$, $V_4$ and $V_6$ are also encoded based on the temporal prediction, but their first frames are encoded using the inter-view prediction as well. In the odd views $V_1$, $V_3$ and $V_5$, both the temporal and inter-view predictions are jointly used to improve the compression performance. For instance, to reconstruct a B-view $V_3$, the decoder should access the adjacent P-views $V_2$ and $V_4$ for the inter-view prediction. Therefore, they should be decoded in the order of $V_2$, $V_4$ and $V_3$. The decoding order is important, since it also determines the available information for the concealment of a lost block. The hierarchical B prediction increases coding efficiency, but it causes severe error propagation. Suppose frame T4 in view V2 is lost during the transmission. The loss propagates to the frames in views V1; V2, and V3.

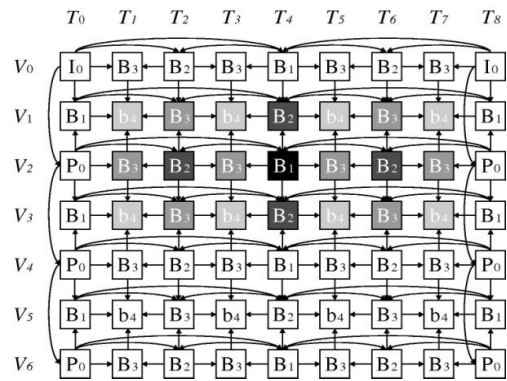

Fig. 5. Hierarchical B prediction structure for 7 view sequence [41].

Considering Motion Vectors are useful for better concealment especially in highly dynamic scenes. In [42], the authors calculate a global DV, for the inter-view referenced frame, relative to the base view and this is transmitted with the anchor frames. When a frame is lost, the corresponding MBs in the dependent frame are located using the global DV. In [20] the authors consider a method that identifies the corresponding region in the reference frame through feature points. This region is used jointly with the boundary pixels in a weighted sum to obtain the replacing MB. In [43], the authors calculate the lost MBs by estimating the MVs and DVs from the neighbourhood temporal and inter-view frames, correspondingly. The outer boundary of the lost MB is considered and a full search for the replacing MB is searched in the temporal and the view-point frames using the Decoder Motion Vector Estimation technique. Other works, such as [44] [45] [46], uses Forward Error Correction (FEC) schemes to introduce redundancies in the codewords to ease their correction. In [44], MBs are classified into slice groups by examining their relative significance to the video and more important MBs are transmitted with better protection, by using the explicit type FMO [45]. In [46] the unequal error protection is formed by defending different frame types with different levels of protection. Intra coded frames are highly protected, followed by temporal predicted frames and inter-view predicted frame. Although this provides good error resilience, it increases the transmission bandwidth. With these schemes simple error concealment techniques can be performed.

## IV. SUBJECTIVE TESTING

### A. Perceptual Characteristics for Testing

Subjective testing is an important component for evaluating the user's Quality of Experience (QoE). In this testing, a group of human subjects are asked to judge the quality of the video sequence under predefined system conditions. The scores given by observers are averaged to produce the *Mean Opinion Score (MOS) and Confidence Intervals (CI)* [2]. Several perceptual characteristics affect the 3D video QoE including the visual quality, depth quality, naturalness, and visual comfort [47]. The visual quality refers here to the perceived spatio-temporal visual quality of the video, which is a main component for both 2D and 3D video. According to ITU-R BT.2021 [47],

*Video quality* refers the perceived quality of the video provided by the system. This is a main determinant of the performance of any video system. Video quality is mainly affected by technical parameters and errors introduced by, for example, encoding and/or transmission processes. *Depth quality* refers to the sensation of depth. While features in monocular cues, such as linear perspective and blur can provide some sensation of depth, stereoscopic 3D images/videos contain extra depth information, which may lead to an enhanced sensation of depth. *Visual (dis)comfort* refers to the degree of comfort when viewing 3D video. Improperly captured, artifacts due to compression, transmission errors, and improperly displayed stereoscopic images could be a source of discomfort. *Naturalness* refers to the degree of the truthful representation of reality for perceived 3D video. It is found that the judgments of naturalness can be split into 75% based on the perceived 2D video quality and 25% based on the perceived depth [47]. *Sense of presence* refers to "the subjective experience of being in one place or environment even when one is situated in another" [48].

### B. Subjective Testingconditions and Methods

For subjective testing of 3D Videos three groups of conditions [49] with source camera sequence can be taken into consideration: (a) Uncompressed and encoded 2D video in full resolution and anamorphic, (b) Uncompressed conditions with different levels of 3D quality and (c) Compressed conditions encoded in Side-by-Side format at different bit rates (constant bit rate encoding). Also, the source 3D sequences can be (a) still camera and small amount of motion, (b) still camera and moderate amount of motion, and (c) zoom, moving/handheld camera, contain from moderate to large amount of motion etc. According to recommendation ITU-R BT.500 [50], the following methods have been successfully used in the last two decades to address relevant research issues related to the picture quality, depth quality and visual comfort of stereoscopic imaging technologies. The methods are single-stimulus (SS), double stimulus continuous quality scale (DSCQS), stimulus-comparison (SC) method and single stimulus continuous quality evaluation (SSCQE) method.

### C. Subjective Evaluation of Concealed 3D Videos

Subjective assessment is commonly used to measure users' quality of experience. The video sequences for the subjective experiments are prepared in a simulated transmission chain, as shown in Figure 6. Several different scenarios, called Hypothetical Reference Circuits (HRC) according to the terminology of the VQEG [51] were used in creating the Video Sequences.

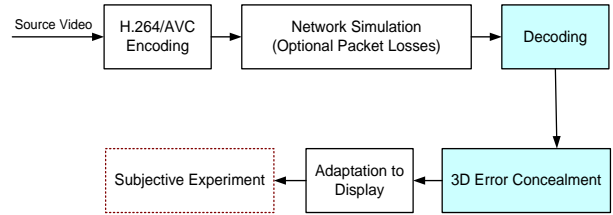

Fig. 6. The processing steps to evaluate the concealed 3D videos

Many standards exist, and they have been used over the years in small and large scale 2D evaluations, e.g. by the Video Quality Experts Group (VQEG) but in 3DTV, some quality parameters such as perceived sharpness or perceived resolution could affect the users' experience. Some new factors in 3D video to the perceived video quality can be expressed in terms of sense of presence and naturalness [52]. All along the transmission chain, the disparity information can be considered [53]. It has also been shown that the visual attention may change when disparity information is available and attention information may be beneficial throughout the transmission chain [54]. The Absolute Category Rating with Hidden unimpaired Reference video (ACR-HR) assessment methods are normally used for subjective testing. As described in VQEG test plan [51]: "ACR is a single-stimulus method in which a processed video sequence is presented on its own, without being paired with its unprocessed "reference" version. Each test condition is randomly presented once to each viewer. The ACR-HR test method includes the non-distorted reference version of each video sequence in order to allow judging the quality of the content itself." In addition to answering on a general five-point ACR scale, the subjects were asked to indicate visual comfort on other scales. Also, there are different time-patterns are being used for testing for adaptation of the viewer as shown in figure 7 [55].

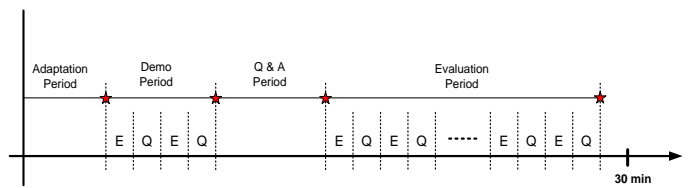

Fig. 7. Time pattern of a subjective evaluation section [55].

## V. PROPOSED ERROR CONCEALMENT

In section III, we have discussed some of the error concealment strategies which are already have studied so far for subjective testing. But recently, the 3D videos achieve more depth perception and viewing for its disparity. So, there is a trade off between disparity and 3D depth. So, if there is an error in a frame, switching to 2D creates high transitional distortion to the observer that will create more visual fatigue and discomfort. Not only that, frame freezing on single or double view create good outcome expect moderate or, high motion videos. And for long duration videos reduced playback speed is more annoying than others. It also creates transitional distortion in high motion videos. To overcome the boundaries, we proposed a low complexity method for full frame loss concealment. It assumed that a buffer of video frame contains one fourth of decoded content which is 6 frames. The 6 frames are played back and forth during the recovery time of the decoder. So, if error occurred in frames and propagates, to make the decoder recover from the errors the last received or stored six frames are used for concealment. Firstly, they are played reverse order and then forward order to make the temporal distortion minimized. It will create less distortion than reduced playback speed and others also.

To compare our method to other existing approaches, we have used five different stereoscopic sequences from RMIT3DV [55], IRCCyN [56] and EPFL [1] have be studies and analyzed. All sequences have duration of 10s and HD resolution 1920*1080@25HZNo audio tracks have been used in the tests and the videos consist of different features regarding the pictorial contents like; camera movement, object motion, texture and sense of presence.

## VI. EXPERIMENTAL RESULTS

The subjective assessment experiments are conducted in a test room confirming the viewing condition defined by BT.500. Foe setting up the experimental conditions, 32 inches Samsung 3D TV has been used. The display uses ACTIVE shutter glassed from the Nvidia 3D vision system. According to VQEG HDTV test plan the viewing distance was 3 times than the display height. According to ITU-R BT.500-11, the display was positioned far from the wall to avoid conflict, no flickering of background light and the illumination is adjusted that no background more than above 15% of the display illumination is considered. The video sequences are played in HD format.

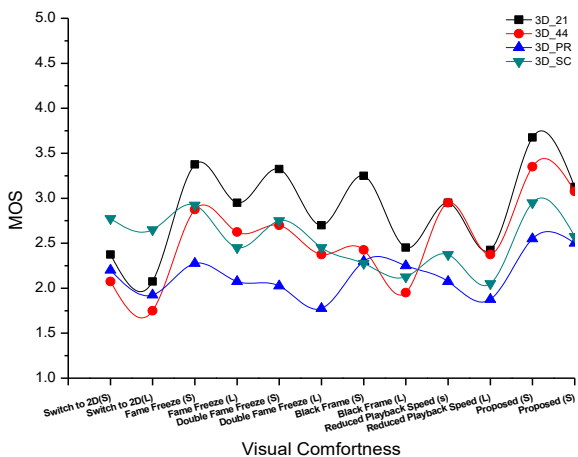

Fig. 8. Comparison of four video sequences. Where MOS is for the Comfortness of different concealment strategies

The concealment techniques have been applied in two types of sequences, which are short (1-5 frames) and long (5-10 frames) durations. Total 10 naïve observers have been participated in the testing experiments. After the experiment, all observers' votes were screened according to ITU-R BT.500 and the VQEG HDTV test plan. For the training session, Double Stimulus Continuous Quality Scale (DSCQS) method was used and for the rating session, Absolute Category Rating with Hidden Reference (ACR-HR) method was used. The training session has been pre-conducted before the formal evaluation session so that observers would familiarize with the rating interface. According to ITU-R BT.2021, two types of perceptual assessments was taken into consideration; Comfortness and Distortion (Sense of Presence). So, for each type of concealment of videos observers have given to ratings for Visual Comfort and Distortion. In Fig. 8 and Fig. 9, we can see the Mean Opinion Score for different types of videos 3D_21, 3d_44, 3d_PR and 3D_SC. The videos contain different types of camera motion, background motion and depth. Through the figure we can find that the proposed method is very much reliable either in motion less or high motion videos than the other concealment strategies.

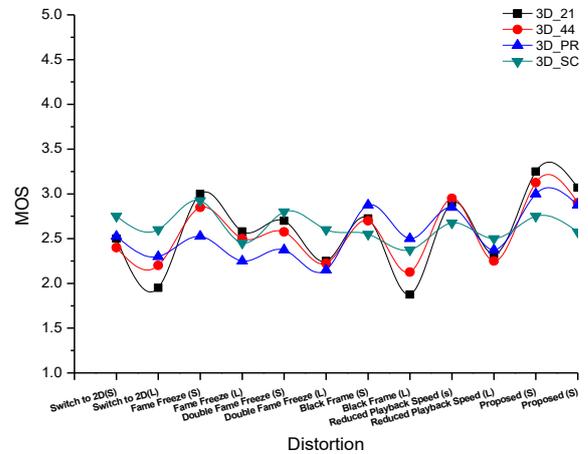

Fig. 9. Comparison of the concealment of four video sequences in respect of visual distortion or sense of presence.

The scores for visual comfort is categorized as 1) extremely uncomfortable, 2) uncomfortable, 3) mildly uncomfortable, 4) comfortable and 5) Very comfortable. For distortion the categorization was 1) worse, 2) bad, 3) fair, 4) good and 5) excellent. By observing the graphs, we can have found that our method gets more scores in terms of comfort and distortion.

## VII. CONCLUSION

The 3D technology and display which will succeed in the future depends crucially upon performance. Above all, they will be judged how realistically natural viewing can be performed after error resilient transmission and proper error concealment. One important step along these lines is to establish subjective assessment and standardized evaluation criteria, which are universally applicable for all types of 3D displays.